# Optical Stark Effect of a Single Defect on TiO$_2$(110) Surface

Lihuan Sun[1,2], Anning Dong[1,2], Jianmei Li[1,2], Dong Hao[1,2], Xiangqian Tang[1,2], Shichao Yan[3], Yang Guo[1,2], Xinyan Shan[1,2*], Xinghua Lu[1,2,4*]

[1]Beijing National Laboratory for Condensed Matter Physics, Institute of Physics, Chinese Academy of Sciences, Beijing 100190, China

[2]School of Physical Sciences, University of Chinese Academy of Sciences, Beijing 100190, China

[3]School of Physical Science and Technology, ShanghaiTech University, Shanghai 201210, China

[4]Center for Excellence in Topological Quantum Computation, Chinese Academy of Sciences, Beijing 100190, China

Probing optical Stark effect at the single-molecule or atomic scale is crucial for understanding many photo-induced chemical and physical processes on surfaces. Here we report a study about optical Stark effect of single atomic defects on TiO$_2$(110) surface with photo-assisted scanning tunneling spectroscopy. When a laser is coupled into the tunneling junction, the mid-gap state of OH-O$_2$ defects changes remarkably in the differential conductance spectra. As laser power gradually increases, the energy of the mid-gap state shifts away from the Fermi level with increase in intensity and broadening of peak width. The observation can be explained as optical Stark effect with the Autler-Townes formula. This large optical Stark effect is due to the tip-enhancement and the strong dipole moment in the transient charged state during electron tunneling. Our study provides new aspects in exploring electron-photon interactions at the microscopic scale.

# I. INTRODUCTION

One of the known phenomena due to light-matter interactions is the optical Stark effect [1], in which a detuned laser excites non-coherent transition of electronic states and forms a series of photon-dressed states [2]. The optical Stark effect is essential for quantum state manipulation and novel optoelectronics applications [3-5], and it has been extensively explored in many structures and materials including quantum dots [6], quantum wells [7, 8] and transition metal dichalcogenides [9-12]. However, most studies, to the best of our knowledge, are carried out based on macroscopic measurements and there has been lack of report on microscopic investigation of local optical Stark effect at the single-molecule or atomic level. With recent progresses in photo-assisted scanning tunneling spectroscopy, it now becomes feasible to probe the local electronic states modulated by the light-matter interactions [13-15]. It is thus of pressing need and fundamental interest to recognize microscopic species on surfaces possessing optical Stark effect and to understand photon-electron interactions at the atomic scale.

In this letter, we report a photo-assisted scanning tunneling microscopy (STM) study of local defects on $TiO_2$(110) surface. $TiO_2$ is a versatile material for various applications such as solar energy harvest [16, 17], photo-catalysis [18, 19], and environment protection [20-22]. The functionality of $TiO_2$, especially its photo-catalysis and photoelectric properties, is closely related to the interactions between light and rich type of defects on the surface [23, 24]. The defect states confined on semiconducting surfaces favor their quantum nature [25] and can be imaged and studied individually by STM [26-29]. Differential conductance (d$I$/d$V$) spectrum taken on a specific kind of local defect, OH-$O_2$, presents remarkable photon-induced changes in its local electronic state, including energy position shift, increase in the intensity and broadening of peak width. The observation reveals a local optical Stark effect that can be explained with the Autler-Townes formula. The energy shift of the local state is on the order of meV, which is due to the tip-enhancement effect and strong dipole moment in the transient charged state during tunneling.

## II. EXPERIMENTAL

The experiment is carried out in a homebuilt low-temperature photo-assisted scanning tunneling microscope [30]. The pressure in the ultra-high vacuum chamber is below $1\times10^{-10}$ torr. Electrochemically etched tungsten tips were used in the experiment. The rutile $TiO_2(110)$ sample crystal was bought from MTI company. The sample was first reduced by hours of annealing at 1100 K until its color changed from transparent to blue. Oxygen vacancies were created in the bulk, resulting in an increase in conductivity as required by low temperature STM measurements. Ordered atomic terraces were then obtained by cycles of Argon ion sputter and anneal. The pulsed laser was generated by a Ti: sapphire oscillator laser (Coherent, Chameleon Ultra II), with 80 MHz repetition rate and 140 fs pulse duration. The laser beam was focused onto STM junction, with a spot size of about 20 µm. The experimental setup is shown in figure 1(a). The wavelength of laser used in this experiment is 810 nm, equivalent to 1.53 eV in photon energy. Laser pulses are coupled with the tunneling junction at an incidence angle of about 45°, with its electric field polarized parallel to the tip. $dI/dV$ spectrum was taken using the lock-in technique with a bias modulation frequency of 264 Hz and amplitude of 10 mV. The temperature of the measurements was around 18 K.

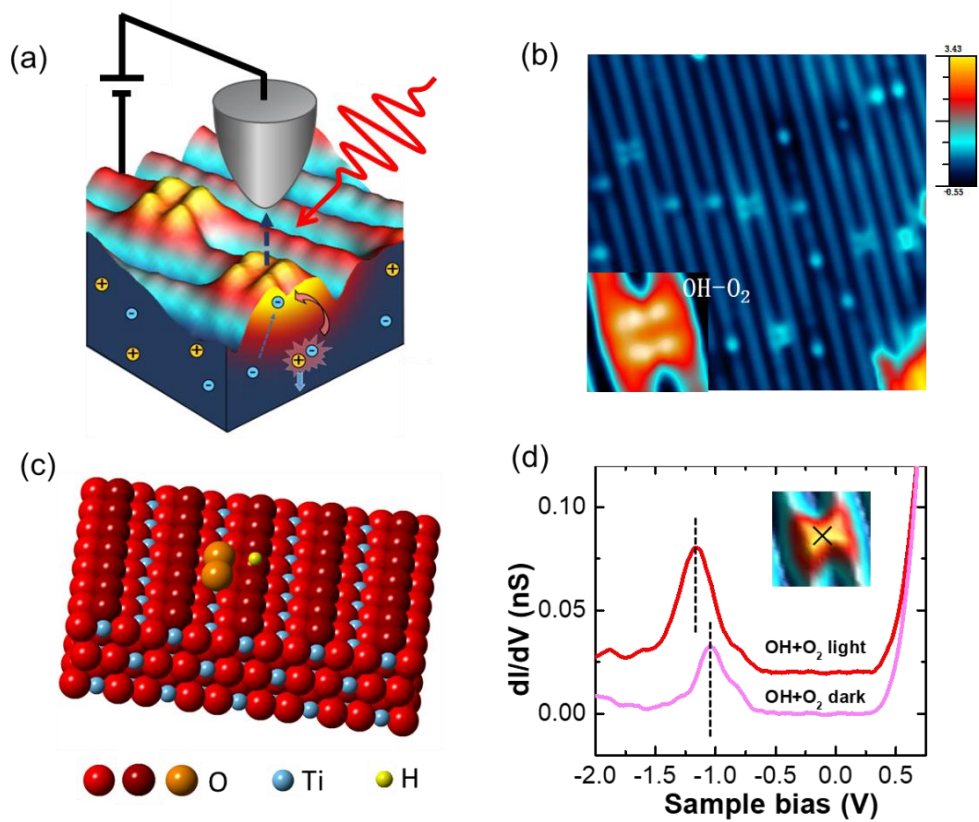

Figure 1. (a) Schematic diagram for the experimental setup. (b) Constant current STM topographic image taken on TiO$_2$(110) surface, taken with sample bias V$_B$ = 1.2 V, I$_T$ = 0.1 nA. Image size: 8.2 nm×8.2 nm. The dark and bright rows correspond to chains of Ti$_{5c}$ and bridge-bonded oxygen along the [001] direction on the surface, respectively. The inset is a detailed image of defect OH-O$_2$. Image size: 1 nm×1 nm. (c) Atomic model of an OH-O$_2$ defect on TiO$_2$(110) surface. (d) d$I$/d$V$ spectra taken on an OH-O$_2$ defect (position indicated by cross in the inset) with and without laser illumination. The spectra are shifted vertically for clarity.

When the sample was initially cooled down to liquid Helium temperature, lots of adsorbed water molecules and hydroxyl groups were often seen on the surface [31]. As a common practice, scanning with high bias voltage (normally >3V) was employed to wipe off most of hydroxyl and water molecules and to get a less defective surface [26, 32]. A typical topographic image of the surface after such operation is shown in Figure 1(b). The bright Ti chain and dark bridge oxygen chain are clearly resolved on the atomic flat terraces. Thanks to previous studies [28, 31, 33, 34], various defect species on the surface can be readily identified, including hydroxyl (OH), bridge-bond oxygen vacancy (BBO$_v$), oxygen atom (O$_{Ti}$), and (OH-O$_2$) [35]. The OH and BBO$_v$ both sit directly on the oxygen chain, while the OH has a relatively higher profile. The $O_{Ti}$ locates directly on the Ti chain, and the OH-$O_2$ presents an asymmetric "x" shape across two Ti chains, as shown in the inset of Figure 1(b). Topographic images of OH-$O_2$ by different bias voltages are consistent with previous experimental reports, and the atomic model of this defect is shown in Figure 1(c) [36].

To explore the electron-photon interaction on $TiO_2$ surface, a pulsed laser beam is coupled onto the STM tunneling junction while taking the d$I$/d$V$ spectra. No significant change was observed in the d$I$/d$V$ spectra taken on both bare substrate and most defect species except the OH-$O_2$ defect. Figure 1(d) presents the spectra taken on the center of an OH-$O_2$ defect, with and without laser illumination. With laser illumination on, the center energy of the mid-gap state is shifted away from the Fermi level, and the peak intensity of local density of states (LDOS) is apparently increased.

To investigate the influence of laser power in the mid-gap state of OH-$O_2$, we varied the laser power gradually from 0 to 11.5 mW. Taking the value of laser pulse width of 150 fs and the laser spot size of 20 μm in diameter, the maximum power corresponds to a peak intensity of about 0.31 GW/cm$^2$, or pulse energy flux of about 0.047 mJ/cm$^2$. The d$I$/d$V$ spectra taken at different laser power are shown in Figure 2. As laser power increases, the center energy of the mid-gap state presents a continuous shift up to 0.2 eV, away from the Fermi level, the peak intensity is magnified by 2 fold, and the full width at half maximum (FWHM) is creased from 0.27 eV to 0.42 eV.

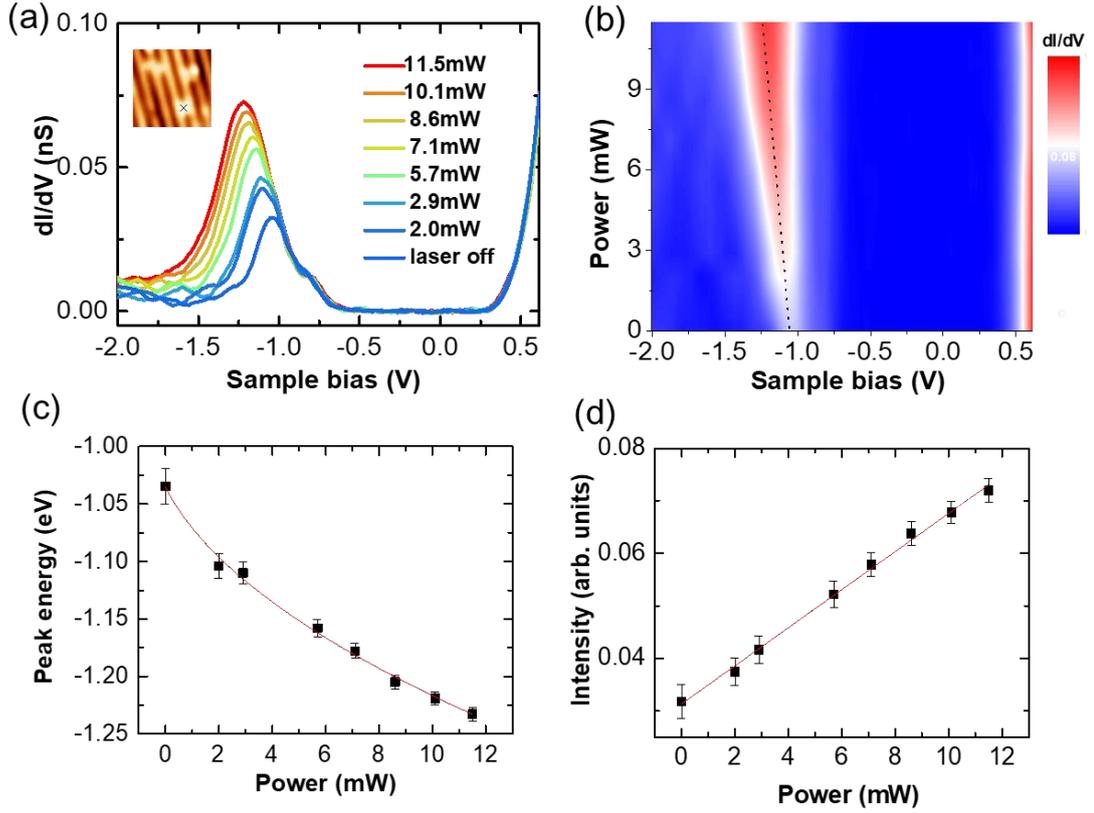

Figure 2. (a) d*I*/d*V* spectra taken on the center of an OH-O$_2$ defect with different laser power from 0 to 11.5 mW. (b) Color map image for the tunneling spectra as a function of sample bias (x axis) and laser power (y axis). (c) Mid-gap state energy as a function of laser power. The fitting curve is based on equation (2). (d) Peak intensity of the mid-gap state as a function of laser power. The red line is the result of a linear fit.

## III. DISCUSSION

Surface photovoltage effect is a common phenomenon associated with band bending on semiconducting surfaces [37-40]. On TiO$_2$ surface, the band bending is very complicated due to the high density of various defects [41-44]. To minimize the photo-voltage effect, we employ low energy photons (1.53 eV) in our experiment. Such an arrangement not only excludes the direct excitation of the carrier, but also minimizes the two-photon effect as even the two-photon energy barely reaches the band gap energy of the substrate. The experimental data shows that there is no energy level shift in photo-assisted d*I*/d*V* spectra taken on bare surface or on most defects except the OH-O$_2$, as shown in Figure 3(a). It is thus evident that the photo voltage effect in our experiment is negligible.

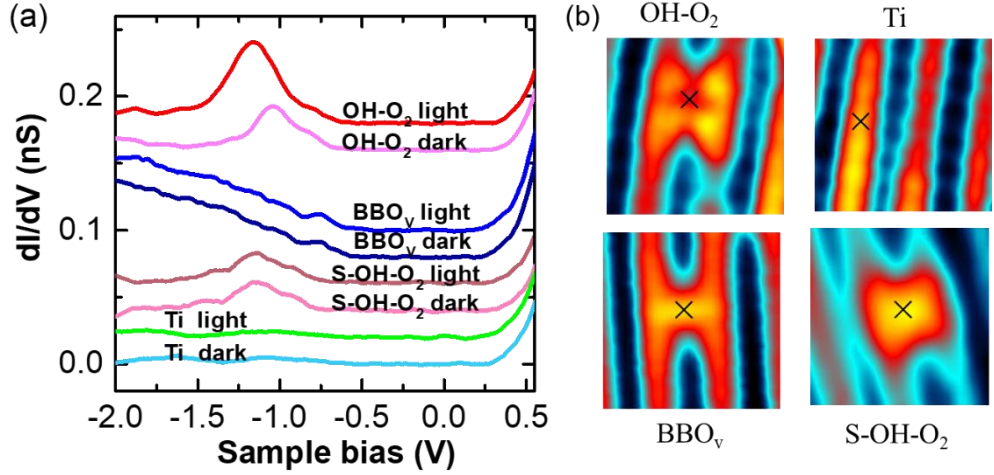

Figure 3: (a) Photo-assisted d$I$/d$V$ spectra taken on various defects and on bare surface. (b) STM images of defects and bare surface where the spectra in (a) were taken. Images are taken with sample bias $V_B$=1.2 V for the top two images and $V_B$=2.0 V for the bottom two images, $I_T$=0.1 nA. Size of the images is 1.5 nm×1.5 nm. S-OH-$O_2$ is a defect similar to OH-$O_2$.

Considering the fact that the OH-$O_2$ defect is a mid-gap state and it has weak interaction with electrons in the bulk of $TiO_2$, the tunneling processes through such a defect can be described by a double-barrier model, as shown in Figure 4(a). In such a model, charging and discharging steps are involved to complete the tunneling process [45]. It has been demonstrated by both theory and experiment that excess electron can be localized on OH-$O_2$ complex [36, 46]. Under zero or small sample bias, the charged state lies above the Fermi level and there is no tunneling current. When a negative sample bias is applied, the charged state moves closer to the Fermi level of the sample due to the voltage divide effect. According to the double-barrier model, the actual energy shift $\Delta E$ is smaller than the observed peak shift in d$I$/d$V$ spectra by a factor of $\frac{1}{1+\varepsilon z/d}$, where $\varepsilon$, $z$ and $d$ are effective dielectric constant of substrate, tip-defect distance, and effective defect-substrate distance, respectively[45]. If we take rough estimated values, $\varepsilon$=6.8, $z$=1 nm and $d$=0.5 nm, the factor is about 0.07. With sufficient negative sample bias, the charged state moves down to the Fermi level and the tunneling current raises. This explains the d$I$/d$V$ spectrum taken on the defect without laser illumination.

When the laser is coupled to the tunneling junction, the energy of the charged state will be shifted due to the electron-photon interaction. Following Autler and Townes's treatment[1], the interaction of the OH-$O_2$ defect and the laser can be described as:

$$H = H_0 + \vec{\mu} \cdot \vec{E} \cos(\omega t), \qquad (1)$$

where $H_0$ is the Hamiltonian of the OH-$O_2$ defect without laser, $\vec{\mu}$ is the dipole of the charged OH-$O_2$ defect, $\vec{E}$ is the effective electric field induced by the laser, and $\omega$ is the laser frequency. By solving the time-dependent Schrodinger's equation, one can get the shift in energy levels and the evolution of the states. The energy shift induced by the laser can be approximately expressed as:

$$\Delta E = \frac{\hbar\omega - E_0}{2} \pm \sqrt{\left(\frac{\hbar\omega - E_0}{2}\right)^2 + \beta_{ab}^2}, \qquad (2)$$

Where $E_0$ is the energy of the excited charged state (relative to the ground state) derived from $H_0$, $\beta_{ab} = \frac{\langle a|\vec{\mu} \cdot \vec{E}|b\rangle}{2}$ is the dipole-field interaction term, the "+" sign is for the excited state and the "−" sign is for the ground state. $\beta_{ab}^2$ is directly proportional to the laser intensity $I$, and can be expressed by $\beta_{ab}^2 = \alpha * I$. We note that the original spectral feature (without laser illumination) of the mid-gap state remains under pulse laser illumination. This implies that the lifetime of the excited state is shorter than but comparable to the repetition period (12.5 ns). To illustrate the new feature resulting from the interaction with light, we tried to subtract the background spectral feature. Simple subtraction, however, results in asymmetric spectral features which is unreasonable. We find that subtract 20% of the background renders the most symmetric spectral features. The peak energy and intensity of new spectra are plotted in Figure 2(c) and (d). In Figure 2(c), we fitted the energy level shift as a function of laser intensity, which derives a value of $2.8\times10^{-5}$ (eV)$^2$/mW for the parameter $\alpha$ and 10 meV for the detuning $\hbar\omega$-$E_0$, after taking the estimated value for the factor $\frac{1}{1+\varepsilon z/d}$.

As shown Figure 2(d), the peak intensity of the mid-gap state changes linearly as increasing the laser power. This can be understood from the evolution of the charged state after the excitation by short laser pulse. The probability of finding the defect in the excited state, at time $t$ after excitation, can be written as

$$|T_b|^2 = \frac{4\beta_{ab}^2}{(\hbar\omega - E_0)^2 + 4\beta_{ab}^2} \sin^2\left(\frac{\sqrt{(\hbar\omega - E_0)^2 + 4\beta_{ab}^2}}{2\hbar} t\right). \quad (3)$$

Since the oscillation period is much longer than the laser pulse duration ($\tau=150 fs$), the expression can be approximated to $|T_b|^2 = \beta_{ab}^2 \tau^2/\hbar^2 = \alpha\tau^2 I$. The tunneling current is thus increased due to the additional tunneling channel through the photo-excited charged state. Such contribution is expected to be proportional to the laser intensity, as demonstrated by the linear fit in Figure 2(d).

We have also noticed the broadening of the spectral feature. The full width at half maximum (FWHM) of the peak is increased from 0.27 eV to 0.42 eV as laser intensity is increased to 11.5 mW. There are several possible reasons for the broadening of the defect-related spectral feature, such as temperature effect and excitation-induced dephasing (EID). Taking into account the factor $\frac{1}{1+\varepsilon z/d}$, the broadening of the peak is about 10 meV, which can be a result of a local temperature up to 120 K. The EID is a nonlinear phenomenon in strong field-electron interacting systems that has been reported for several semiconducting quantum dots [47-49], where the spectra linewidth increases with the excitation power. Unfortunately, present experimental data is not adequate to explain the reason for broadening yet, and further investigations are desired.

From the peak energy shift of 0.2 eV in d$I$/d$V$ spectra, we can estimate the energy shift in the charged state as 14 meV, by implementing the factor $\frac{1}{1+\varepsilon z/d}$. This value is larger than but close to the values reported in other systems, such as germanium quantum wells [8], monolayer $WS_2$ [9], and lead-halide perovskites [50]. (Keep in mind that the value of 0.07 for the factor is a rough estimate, its low limit could be 10 times smaller.) This large effect is due to both the tip-enhancement effect and the extraordinary dipole matrix element $\beta_{ab}$. The sharp STM tip can focus the electromagnetic field to the tunneling junction, resulting an enhancement factor about 10 [51-53]. The dipole matrix element is expressed by a spatial integration of the initial ground state and the excited charged state, $\beta_{ab} = \dfrac{\langle a|\vec{\mu}\cdot\vec{E}|b\rangle}{2}$. The final charged state contains an electron on the defect and a hole in the surrounding substrate, which has an electronic dipole moment on the order of $1e\cdot nm$. Apparently, the magnitude of this dipole is significantly larger than the dipoles in free atoms or molecules. The large optical Stark effect is thus rational under the above considerations.

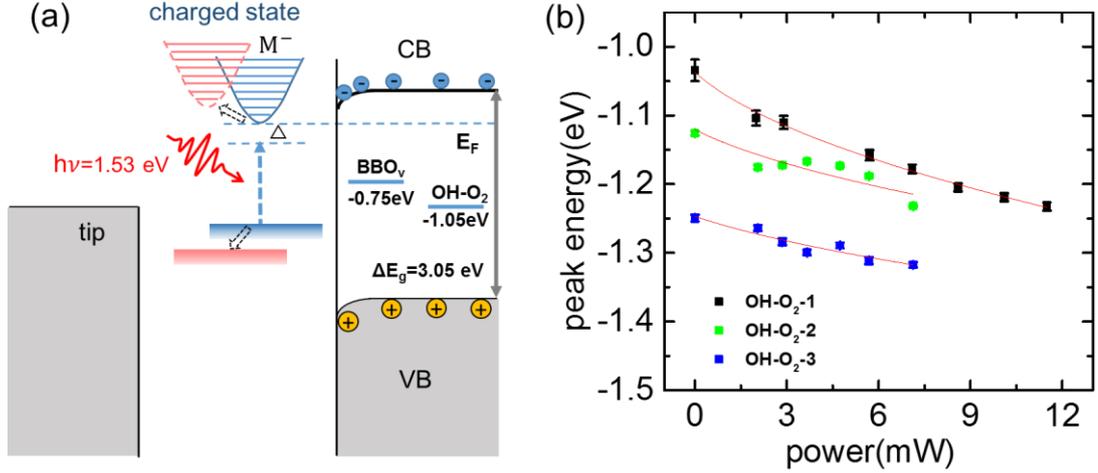

Figure 4. (a) Schematic diagram for optical Stark effect in a local defect on semiconducting surface based on the double-barrier tunneling model. Defect energy levels are referred to the Fermi level. The tunneling process occurs when the charged state aligns with the Fermi level in the substrate. (b) Mid-gap state energy of three OH-$O_2$ defects as a function of laser power.

More interestingly, we find that the optical Stark effect in single defects on surface is very sensitive to the local environment. In Figure 4(b), we present energy shifts in the mid-gap states of three OH-$O_2$ defects. By fitting with Eq. (2), similar values of detuning energy are derived. The value of $\alpha$, however, differs significantly, as of $2.8\times10^{-5}$, $1.5\times10^{-5}$, $1.0\times10^{-5}$ (eV)$^2$/mW for the defects 1, 2, 3, respectively. Defects with deeper energy in the spectra is associated with a stronger band bending when the sample bias was set to the center of the mid-gap states. The stronger the band bending, the more electrons accumulated on surface near the defect, resulting in stronger Coulomb screening effect and thus weaker dipole moments. This explains the decrease in value of $\alpha$ for defects 2 and 3, and it provides further evidence that such optical Stark effect is directly related to the dipole of the charged state. The sensitivity of optical Stark effect on local environment introduces possible applications of gate-controlled photo-electronic devices based on single defects and renders photo-assisted scanning tunneling spectroscopy a unique and irreplaceable experimental tool in exploring this local phenomena.

## IV. SUMMARY

To conclude, we have observed a large optical Stark effect in a single defect on TiO$_2$(110) surface by using photo-assisted scanning tunneling spectroscopy at low temperature. The experimental result is well explained by introducing electron-photon interaction in a two-level quantum system. The large energy shift is due to the tip enhancement and extraordinary electronic dipole moment associated with the transient charge state. The experimental method as demonstrated in this study provides a unique route towards exploring and understanding electron-photon interaction at the single-molecule or atomic scale.

## ACKNOWLEDGMENT

This research is supported by National Science Foundation of China under Grant Nos 11774395 and 91753136, Strategic Priority Research Program (B) of the Chinese Academy of Sciences under Grant No. XDB07030100 and XDPB0601, and Beijing Natural Science Foundation (4181003). The authors thank prof. Jimin Zhao for data discussion.